\documentclass[%
superscriptaddress,
amsmath,amssymb,
aps,
twocolumn,
]{revtex4-2}

\usepackage{hyperref}
\hypersetup{colorlinks,allcolors=blue}
\usepackage{graphicx}
\usepackage{bm}
\usepackage[all]{hypcap} 
\usepackage{color}
\usepackage{siunitx}
\usepackage{mathtools}
\usepackage{ulem}
\begin{document}

\title{Evolution of the electronic structure in Ta$_2$NiSe$_5$ across the structural transition revealed by resonant inelastic x-ray scattering}

\author{Haiyu Lu}
\affiliation{Stanford Institute for Materials and Energy Sciences, SLAC National Accelerator Laboratory, 2575 Sand Hill Road, Menlo Park, California 94025, USA}
\affiliation{Department of Physics, Stanford University, Stanford, California 94305, USA}

\author{Matteo Rossi}
\affiliation{Stanford Institute for Materials and Energy Sciences, SLAC National Accelerator Laboratory, 2575 Sand Hill Road, Menlo Park, California 94025, USA}

\author{Jung-ho Kim}
\affiliation{Advanced Photon Source, Argonne National Laboratory, Argonne, IL 60439, USA}

\author{Hasan Yavas}
\affiliation{Linac Coherent Light Source (LCLS), SLAC National Accelerator Laboratory, Menlo Park, CA 94025, USA}

\author{Ayman Said}
\affiliation{Advanced Photon Source, Argonne National Laboratory, Argonne, IL 60439, USA}

\author{Abhishek Nag}
\affiliation{Diamond Light Source, Harwell Campus, Didcot OX11 0DE, United Kingdom}

\author{Mirian Garcia-Fernandez}
\affiliation{Diamond Light Source, Harwell Campus, Didcot OX11 0DE, United Kingdom}

\author{Stefano Agrestini}
\affiliation{Diamond Light Source, Harwell Campus, Didcot OX11 0DE, United Kingdom}

\author{Kejin Zhou}
\affiliation{Diamond Light Source, Harwell Campus, Didcot OX11 0DE, United Kingdom}

\author{Chunjing Jia}
\affiliation{Stanford Institute for Materials and Energy Sciences, SLAC National Accelerator Laboratory, 2575 Sand Hill Road, Menlo Park, California 94025, USA}

\author{Brian Moritz}
\email{moritzb@slac.stanford.edu}
\affiliation{Stanford Institute for Materials and Energy Sciences, SLAC National Accelerator Laboratory, 2575 Sand Hill Road, Menlo Park, California 94025, USA}

\author{Thomas P. Devereaux}
\affiliation{Stanford Institute for Materials and Energy Sciences, SLAC National Accelerator Laboratory, 2575 Sand Hill Road, Menlo Park, California 94025, USA}
\affiliation{Department of Materials Science and Engineering, Stanford University, Stanford, California 94305, USA}
\affiliation{Geballe Laboratory for Advanced Materials, Stanford University, Stanford, CA, USA}

\author{Zhi-Xun Shen}
\email{zxshen@stanford.edu}
\affiliation{Stanford Institute for Materials and Energy Sciences, SLAC National Accelerator Laboratory, 2575 Sand Hill Road, Menlo Park, California 94025, USA}
\affiliation{Department of Physics, Stanford University, Stanford, California 94305, USA}
\affiliation{Geballe Laboratory for Advanced Materials, Stanford University, Stanford, CA, USA}

\author{Wei-Sheng Lee}
\email{leews@stanford.edu}
\affiliation{Stanford Institute for Materials and Energy Sciences, SLAC National Accelerator Laboratory, 2575 Sand Hill Road, Menlo Park, California 94025, USA}


\begin{abstract}
We utilized high-energy-resolution resonant inelastic X-ray scattering (RIXS) at both the Ta and Ni $L_3$-edges to map out element-specific particle-hole excitations in Ta$_2$NiSe$_5$ across the phase transition. Our results reveal a momentum dependent gap-like feature in the low energy spectrum, which agrees well with the band gap in element-specific joint density of states calculations based on ab initio estimates of the electronic structure in both the low temperature monoclinic and the high temperature orthorhombic structure. Below $T_c$, the RIXS energy-momentum map shows a minimal gap at the Brillouin zone center ($\sim$ 0.16 eV), confirming that Ta$_2$NiSe$_5$ possesses a direct band gap in its low temperature ground state. However, inside the gap, no signature of anticipated collective modes with an energy scale comparable to the gap size can be identified. Upon increasing the temperature to above $T_c$, whereas the gap at the zone center closes, the RIXS map at finite momenta still possesses the gross features of the low temperature map, suggesting a substantial mixing between the Ta and Ni orbits in the conduction and valence bands, which does not change substantially across the phase transition. Our experimental observations and comparison to the theoretical calculations lend further support that the phase transition and the corresponding gap opening in Ta$_2$NiSe$_5$ is largely structural by nature with possible minor contribution from the putative exciton condensate.

\end{abstract}


\maketitle
\section*{Introduction}

The excitonic insulator (EI) is a long-conjectured quantum phase  that could occur in narrow-gap semiconductors or semimetals. In these systems, electron-hole pairs (excitons) can be created spontaneously and Bose condense into a collective state at low temperatures \cite{Mott1961, kozlov1965metal, keldysh1968collective, Jerome1967, Halperin1968}. The condensation gives rise to an insulating ground state, characterized by the opening of an energy gap due to hybridization between conduction and valence bands. Since the EI is an electronic many-body state with a close analogy to superconductivity, it is of great interest to search for materials that can realize the EI state.

Many of the EI candidates, such as 1T-TiSe$_2$ \cite{Cercellier2007, Kogar1314} and TmSe$_{0.45}$Te$_{0.55}$ \cite{Bucher1991, Bronold2006}, also are known to exhibit a charge density wave (CDW) state across the putative EI phase transition. Indeed, since these systems possess an indirect band gap, an EI state inevitably induces a lattice distortion, giving rise to a parasitic CDW state with an ordering wavevector corresponding to the momentum separation between the top of the valence band and the bottom of the conduction band. However, other experiments also suggest that the CDW could be induced by other mechanisms unrelated to the EI \cite{Kidd2002, Van2010}, which has complicated the unambiguous identification of EI in these materials. 

To circumvent the parasitic CDW induced by an indirect band gap, Ta$_2$NiSe$_5$ \cite{Sunshine2006, DiSalvo1986} originally was proposed \cite{Wakisaka2009, Kaneko2012} as a cleaner material that may host an EI phase, due to its direct band gap. Transport and specific heat \cite{Lu2017} measurements have confirmed the existence of a second-order phase transition at $T_c$ = 326 K in Ta$_2$NiSe$_5$. Optical conductivity \cite{Lu2017, Larkin2017}, angle-resolved photoemission spectroscopy (ARPES) \cite{Wakisaka2009, Wakisaka2012, Kaneko2012, Fukutani2019}, and pump-probe time-resolved measurements \cite{Mor2017, Mor2018, Okazaki2018, Werdehauseneaap8652} have found experimental evidence that the low temperature phase is characterized by a strong hybridization between the conduction band with Ta $5d$ character and the valence band with Ni $3d$ character near the Brillouin zone center. Across the transition from high to low temperature, the valence band top flattens and quasi-particle peak sharpens, while a characteristic energy gap of $\sim$ 0.16 eV opens, consistent with the expectations for an EI phase transition.

Unfortunately, while a CDW is absent, Ta$_2$NiSe$_5$ exhibits a structural transition, posing a similar dilemma as other EI candidates. The crystal structure of Ta$_2$NiSe$_5$ undergoes a small shear-like lattice distortion from orthorhombic to monoclinic across the phase transition, concurrent with the gap opening in the electronic structure. It is still heavily debated whether the gap opening and the associated band hybridization are driven by the electronic (\textit{i.e.} the EI transition) or the lattice degree of freedom (\textit{i.e.} structural transition). Some theoretical analysis \cite{Kaneko2013, seki2014, Lee2019} asserted that the small lattice distortion induced by the structural phase transition alone is too small to induce a sufficient band hybridization and gap as large as $\sim$ 0.16 eV, favoring the EI scenario. However, this notion has been challenged by a number of other studies: a polarization-dependent ARPES study \cite{Watson2020} claimed that the lattice distortion by itself can explain the band hybridization and the energetics of the phase transition. Moreover, by evaluating the temporal dependence of the hybridization gap, near-infrared pump time-resolved ARPES and ultrafast electron diffraction (UED) measurements hinted that the phase transition could not be driven exclusively by the electronic degrees of freedom \cite{Tang2020, Baldini2020}. In addition, a Raman experiment reported the softening of an optical zone-center phonon associated with the lattice distortion, indicating a significant involvement of the lattice degree of freedom \cite{Kim2020}. 

To clarify this debate, it would be useful to obtain information about the full gap structure and the associated band hybridization across the phase transition. While ARPES has revealed extensive information about the gap opening associated with the occupied valence bands, the full momentum dependent gap structure is lacking between the occupied valence and unoccupied conduction bands, and element-specific information that could reflect the Ta-Ni hybridization is scarce in available experimental studies. 

\begin{figure}
	\centering
	\includegraphics[width=0.9\columnwidth]{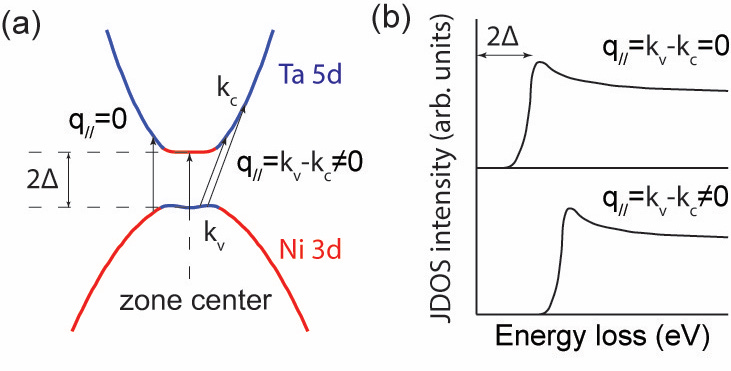}
	\caption{\label{fig1} (a) Schematics of occupied valence and unoccupied conduction bands near the Fermi energy. A many-body gap of $\sim 2\Delta$ is opened due to the formation of a putative EI phase transition at low temperatures. $q_{//} = 0$ and $q_{//} \ne 0$ transitions across the gap are shown by the arrows, representing the JDOS. (b) The corresponding JDOS spectra show a gap-like feature , which represents the gap for the given momentum difference in JDOS. The full momentum dependent gap structure near the Fermi energy can be mapped out by the JDOS.}
\end{figure}

Such information about the full gap between the occupied valence and unoccupied conduction bands is encoded in the momentum-resolved joint density of states (JDOS). Figure~\ref{fig1}(a) shows a cartoon of the valence and conduction bands with a gap $\sim$ $2\Delta$ due to the formation of the putative EI state at low temperatures. Under these circumstances, the JDOS, which is the convolution of the density of states between specific electronic momentum in the valence ($k_v$) and conduction bands ($k_c$), exhibits a gap at the corresponding momentum difference $q_{//}=|\bm{k}_v-\bm{k}_c|$, as shown in the Fig.~\ref{fig1}(b). Thus, if the JDOS can be measured experimentally, the momentum resolved full gap structure can be mapped out. 

Resonant inelastic X-ray scattering (RIXS) has become a powerful tool to measure elementary excitations \cite{Ament2011}.In particular, the RIXS cross-section is sensitive to charge excitations with additional modulation from the RIXS matrix elements and polarization \cite{Jia2016}. Thus, RIXS may be used to probe the particle-hole excitations in momentum space across the band gap between occupied and unoccupied bands near the Fermi energy, whose gap structure closely relates to that of the JDOS. Interestingly, a recent Ni $L_3$-edge RIXS study of Ta$_2$NiSe$_5$ \cite{Monney2020} in the low temperature phase has demonstrated the resemblance of RIXS spectrum to the JDOS between the occupied valence and unoccupied conduction bands. In fact, since Ni $L_3$-edge RIXS specifically involves transition on Ni ions, the spectrum should be related closely to the JDOS projected on the Ni-orbital character. However, theoretical calculations show that the Ta $5d$ orbital character is expected to dominate the lowest unoccupied conduction band, it is of great interest to carry out RIXS measurements at both the Ta and Ni $L$-edges to obtain element-specific information about the electronic structure. Moreover, since RIXS is a charge neutral process, in which the scattered photons do not add or remove charge from the material, RIXS should be sensitive to electronic collective modes of the exciton condensate whose energy scale is expected to be the same order of the EI gap size. The observation of such modes could serve as a direct evidence of the exciton condensate.

In this study, we report results of both hard X-ray RIXS at the Ta $L_3$-edge and soft X-ray RIXS at the Ni $L_3$-edge using state-of-the-art RIXS instruments with high energy resolution. The RIXS spectra reveal momentum dependence of the full gap structure across the phase transition temperature and agree well with the element-specific joint density state deduced from the Ta- and Ni-specific JDOS deduced from ab initio band structure calculations which solely rely on the corresponding lattice structures. However, within our instrument resolution, we do not find any signature of a collective mode related to exciton condensation inside the gap.

\section*{Experiment details}

High quality single crystals of Ta$_2$NiSe$_5$ ($\sim$ 10.0 × 0.5 × 0.1 $mm^3$ in size) were synthesized by chemical vapor transport method \cite{Lu2017}. Elemental powders of tantalum, nickel and selenium (Sigma-Aldrich, purity, 99.99$\%$, 99.99$\%$, 99.99$\%$) were mixed with a stoichiometric ratio and sealed into a quartz tube ($\sim$ 10$^{-3}$ $pa$) with a small amount of iodine as the transport agent. The mixture was reacted under a temperature gradient of 950 to 850 $^{\circ}C$. After 10 days, needle-shaped Ta$_2$NiSe$_5$ single crystals were synthesized at the cooler end of the tube. The crystallinity of the sample was characterized by X-ray diffraction (XRD) at room temperature using Cu-$K_\alpha$ radiation (PANalytical X'Pert PRO X-ray diffraction system). The transport $T_c$ at 326 K was measured using the resistivity module of a Physical Property Measurement System from Quantum Design. The crystal orientation was determined and aligned using Laue diffraction prior to RIXS measurements. 

The hard X-ray RIXS measurements at Ta $L_3$-edge were performed at the Sector 27-ID RIXS beamline of the Advanced Photon Source using the RIXS spectrometer, where the sample temperature could be cooled down to 25 K or heated up to 335 K \cite{Gog2012, Shvydko2013}. X-rays were monochromatized to a bandwidth of 38 meV, and focused to have a beam size of 40 (horizontal) $\times$ 15 (vertical) $\mu m^2$. A Si (066) diced spherical analyzer with a 4 inch radius and a position-sensitive silicon microstrip detector were used in the Rowland geometry. The soft X-ray RIXS measurements at Ni $L_3$-edge were carried out at beamline I21, Diamond Light Source in the UK, where the sample temperature was fixed at 20 K. The combined energy resolution of the monochromator and spectrometer was around 100 meV at the Ta $L_3$-edge and 40 meV at the Ni $L_3$-edge, as determined from the full-width-half-maximum of the elastic peak. X-ray absorption spectroscopy (XAS) spectra were measured before collecting RIXS data. The RIXS data at both the Ta and Ni $L_3$-edge were collected with linear horizontal polarization (parallel to the scattering plane, $\pi$-polarization) of the incident X-ray beam. 

\begin{figure*}
	\centering
	\includegraphics[width=2.0\columnwidth]{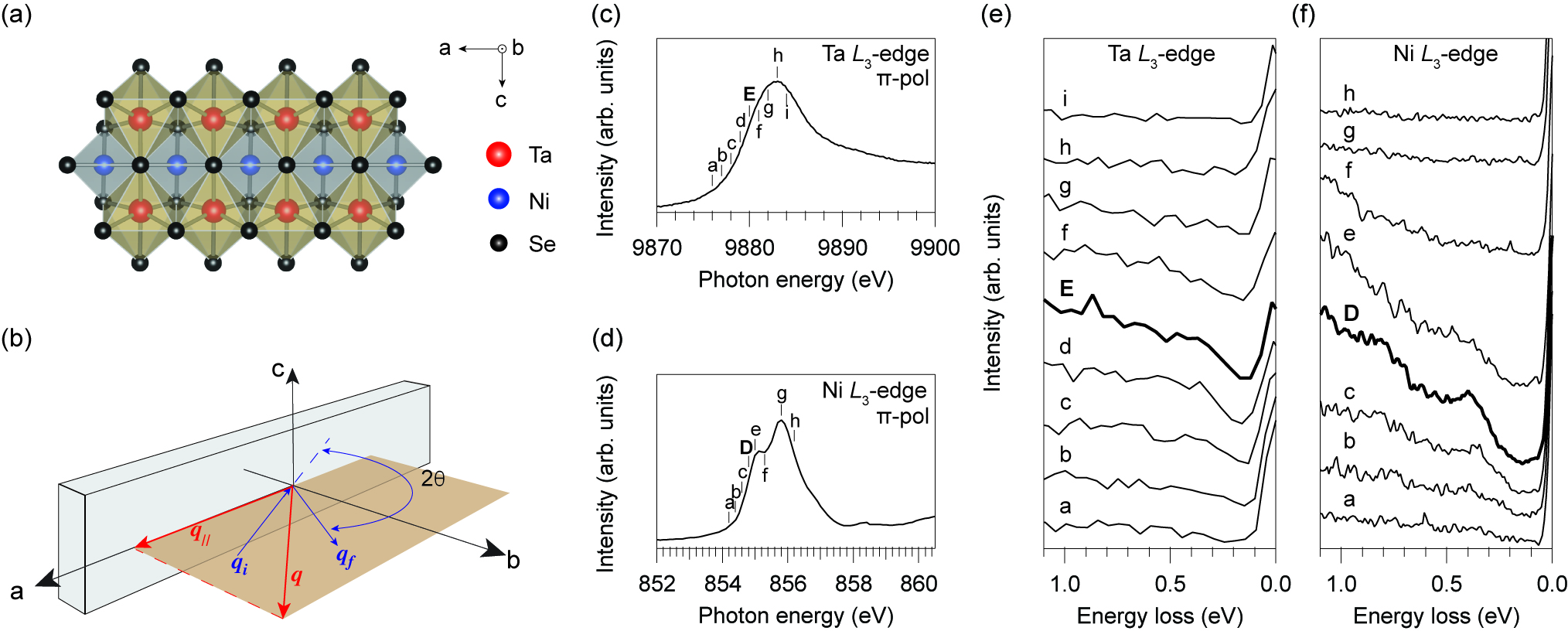}
	\caption{\label{fig2} (a) Crystal structure of needle-shaped Ta$_2$NiSe$_5$. (b) Schematics of RIXS scattering geometry. We aligned the Ta$_2$NiSe$_5$ sample such that the chain direction ($a$ axis) is in the RIXS scattering plane. (c) and (d) Ta and Ni $L_3$-edge XAS spectra measured with $\pi$-polarized incident X-ray light in the grazing and normal incidence geometry, respectively. (e) and (f) Ta/Ni $L_3$-edge incident energy dependence RIXS spectra measured across the corresponding XAS resonant peaks at 25/20 K with momentum transfer along the chain equal to 0 and 0.025 $r.l.u.$, respectively. The RIXS gap feature is most pronounced at the XAS pre-peak position (labelled by the bold/capital letter) for both the Ta and Ni $L_3$-edge. The corresponding RIXS spectra are indicated by thicker lines.}
\end{figure*}

Ta$_2$NiSe$_5$ is a layered compound stacked by van der Waals interaction with parallel chains of Ta and Ni atoms (See Fig.~\ref{fig2}(a)). It is expected that the compound exhibits a quasi-one-dimensional electronic structure along the $a$ axis with a weak out-of-plane dispersion along the $b$ axis. Thus, we mounted the sample such that the chain direction ($a$ axis) was aligned in the scattering plane (See Fig.~\ref{fig2}(b)). In this geometry, the momentum transfer $\bm{q}$ of the $\pi$-polarized incident X-ray beam has a parallel component $q_{//}$ along the $a$ axis (Ta and Ni chain direction) and a perpendicular component $q_{\perp}$ along the $b$ axis. The momentum dependence of the RIXS spectrum was obtained by rotating the sample angle to vary $q_{//}$ while keeping the spectrometer angle the same.

\begin{figure*}
	\centering
	\includegraphics[width=1.6\columnwidth]{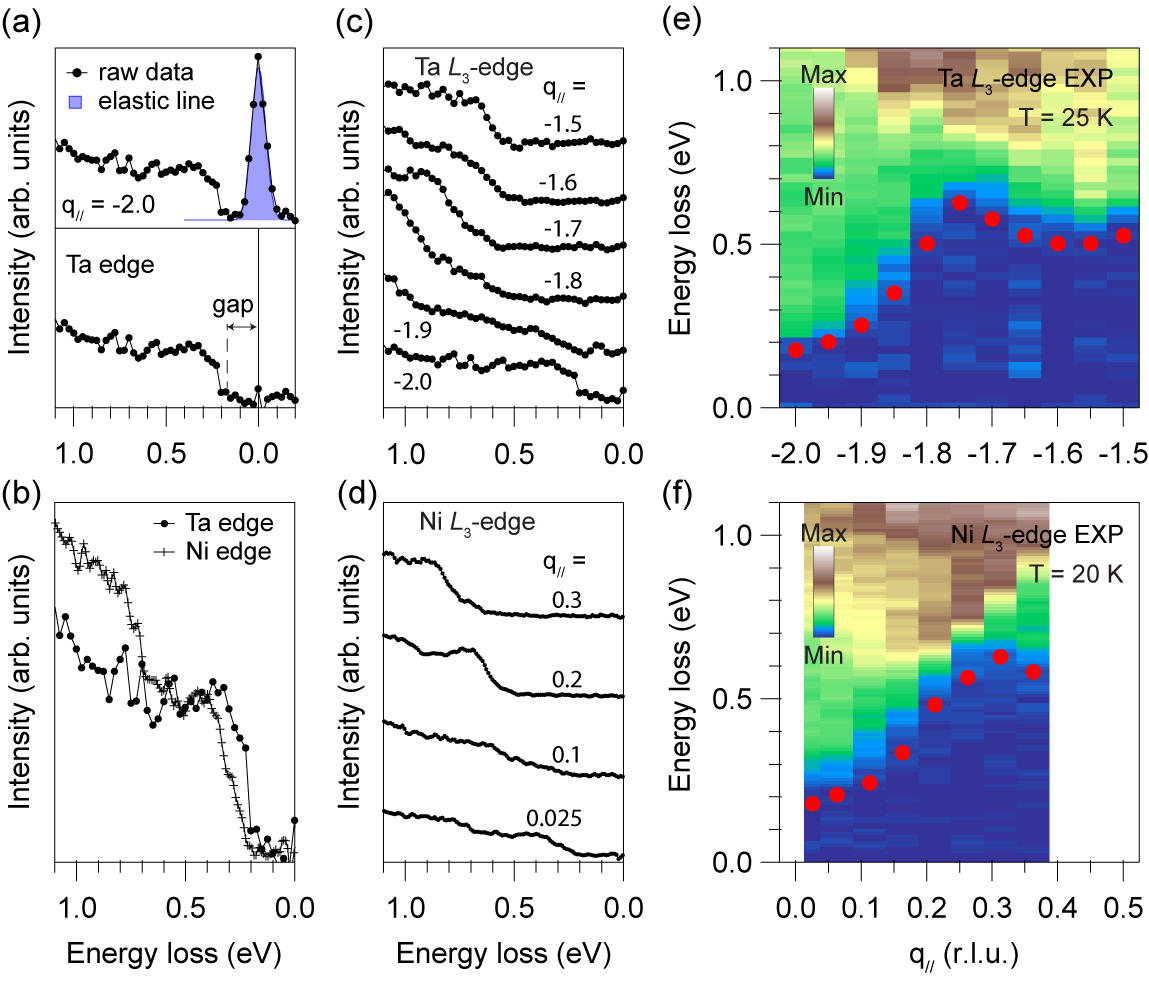}
	\caption{\label{fig3} (a) The quasi-elastic region is subtracted by fitting it to a Gaussian function. The remaining raw data shows a RIXS gap feature. The gap leading edge is well defined by the gap onset. (b) Comparison of Ta/Ni $L_3$-edge RIXS spectra at the zone center. While the gap onsets are identical, the gap leading edges are different between the Ta and Ni $L_3$-edge spectra. (c) and (d) Waterfall plots of quasi-elastic subtracted Ta/Ni $L_3$-edge RIXS spectra at representative in-plane momentum positions from the zone center towards the zone boundary. (e) and (f) Quasi-elastic subtracted color map of Ta/Ni $L_3$-edge momentum dependent RIXS data. The superimposed red dots trace the onset of the RIXS gap feature.}
	
\end{figure*}

\begin{figure*}
	\centering
	\includegraphics[width=1.75\columnwidth]{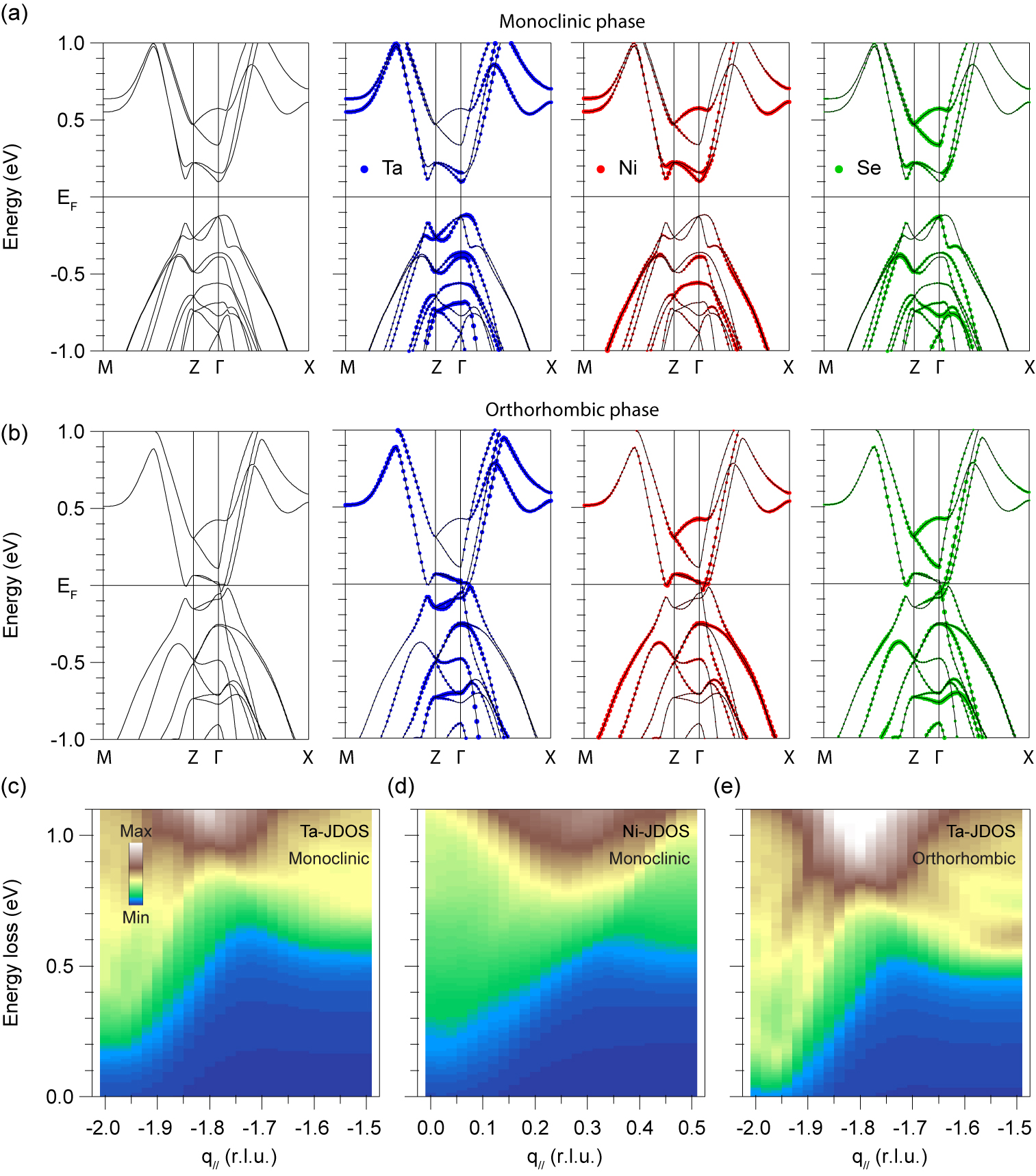}
	\caption{\label{fig4} (a) and (b) DFT electronic bands involved in the element-specific JDOS calculation for the (a) low temperature monoclinic and the (b) high temperature orthorhombic structure. For both low and high temperature DFT bands, the corresponding partial density of states of Ta 5$d$ (blue), Ni 3$d$ (red), and Se 4$p$ (green) are showed. (c) to (e) element-specific JDOS calculations of (c) Ta-JDOS in the low temperature monoclinic phase, (d) Ni-JDOS in the low temperature monoclinic phase, and (e) Ta-JDOS in the high temperature orthorhombic phase. The calculations are in good agreement with the RIXS data and reproduce the subtle element-specific differences between the Ta- and Ni-RIXS intensity maps.}
\end{figure*}

\begin{figure}
	\centering
	\includegraphics[width=0.9\columnwidth]{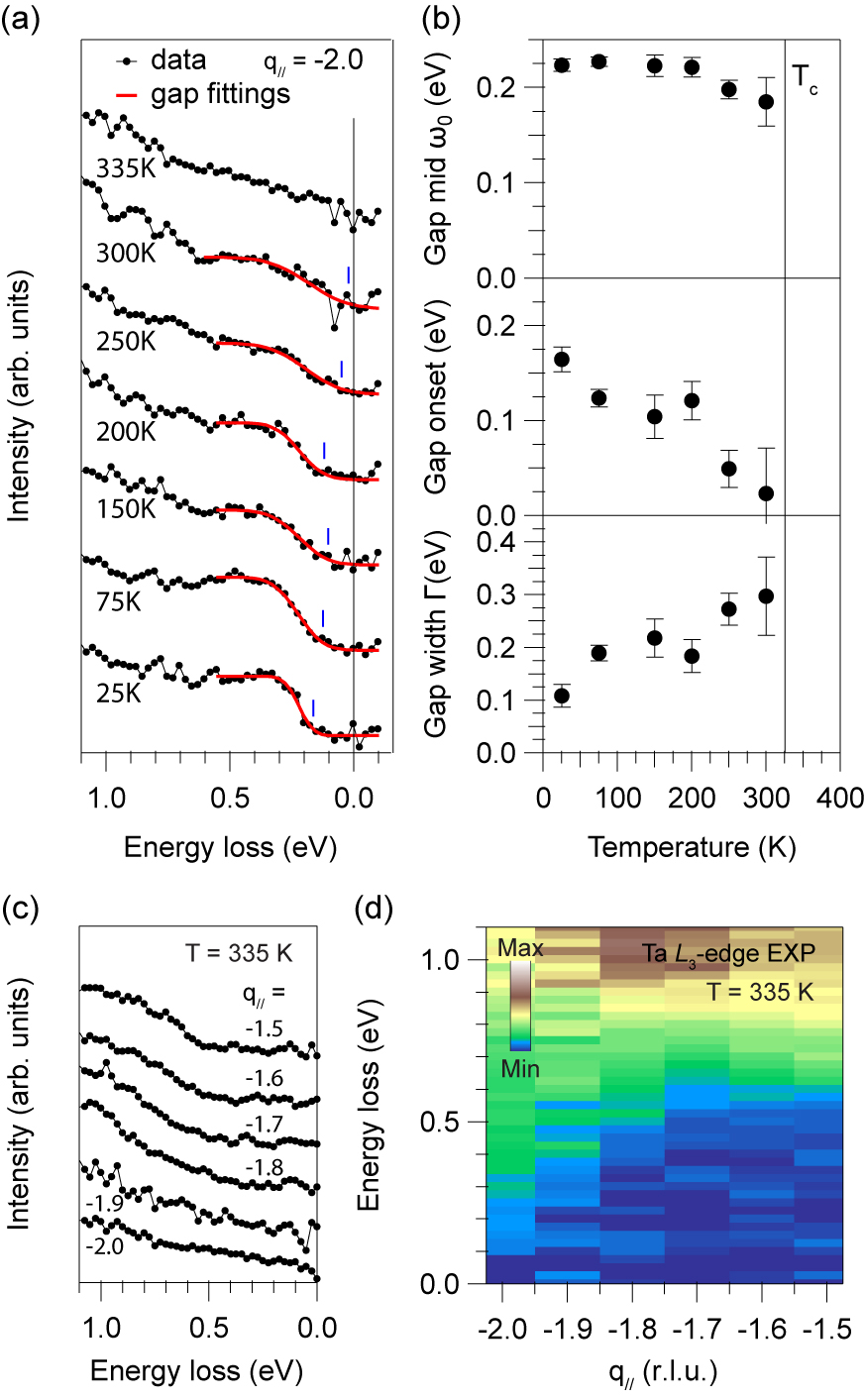}
	\caption{\label{fig5} (a) Quasi-elastic subtracted momentum dependent Ta $L_3$-edge RIXS spectra from the lowest temperature of 25 K to 335K ($T_c$ = 326 K). The RIXS gap gradually closes and gap edge broadens as a function of increasing temperature. The gap fittings are plotted as the red curves. Energy positions of 10$\%$ step height are indicated by the blue ticks. The gap onset is defined by the 10$\%$ ticks and the gap width is defined by the fitting parameter $\Gamma$. (b) A summary of the RIXS gap mid-point energy $\omega_0$, onset and width $\Gamma$ as a function of temperature. The error bars represent the standard deviation of the fitting. The RIXS gap above $T_c$ cannot be fitted reliably within the energy resolution. (c) Waterfall plots of quasi-elastic subtracted Ta $L_3$-edge RIXS spectra at representative in-plane momentum positions from the zone center towards the zone boundary at 335 K. (d) Quasi-elastic subtracted color map of Ta $L_3$-edge momentum dependent RIXS data at 335 K.}
\end{figure}

\section*{Results}

Figure~\ref{fig2}(c) and (d) show the X-ray absorption spectrum of Ta$_2$NiSe$_5$ across the Ta and Ni $L_3$-edge, respectively, measured by total fluorescence yield at temperatures well below $T_c$. The Ta $L_3$-edge XAS shows a broad single-peak over a wide range of 30 eV, while the Ni $L_3$-edge XAS exhibits two main peaks separated by 0.7 eV over a much narrower range of 5 eV. In order to find the optimal incident energy that yields the strongest intensity of the RIXS signal, we conducted incident energy dependence measurements across the XAS edges, as summarized in Fig.~\ref{fig2}(e) and (f). Besides a strong elastic line at 0 eV, a gap-like feature in the RIXS spectrum can be resolved at both the Ta and Ni $L_3$-edge, which was found to be most pronounced when the incident photon energy was tuned to an energy slightly less than the maximum in the XAS.

Next, we fixed the incident photon energy here, denoted E and D in Fig.~\ref{fig2}(c) and (d), respectively, and explored the variation of RIXS spectra as a function of $q_{//}$. At the Ta $L_3$-edge, within the momentum transfer of X-ray, we found that the gap-like feature in the spectrum was most pronounced in the 2$^{nd}$ in-plane Brillouin zone where the equivalent zone center is at $h = -2.0~r.l.u.$ (grazing incidence geometry). As shown in Fig.~\ref{fig3}(a), the quasi-elastic region of the RIXS spectrum can be well represented by a symmetric Gaussian function, so we fit and subtract it from all raw data. Hereafter, we will only focus on the the elastic-peak-subtracted RIXS spectra, unless indicated otherwise. 

A comparison of the Ta and Ni $L_3$-edge RIXS spectra at the zone center is plotted in Fig.~\ref{fig3}(b) and the waterfall plots of RIXS spectra at representative momentum positions are shown in Fig.~\ref{fig3}(c) and (d). As can be seen, the spectra represent a continuum of excitations with an abrupt onset of the leading edge, suggesting the existence of an energy gap in the underlying particle-hole excitations. In addition, at both the Ta and Ni $L_3$-edge, the RIXS gap at the zone center onsets at approximately 0.16 eV, which is consistent with the energy gap in the density of states as measured by optical conductivity \cite{Lu2017, Larkin2017} and scanning tunneling spectroscopy (STS) \cite{Lee2019}. By examining the momentum dependence of the gap, as shown in Fig.~\ref{fig3}(c)-(f), we find that the gap is minimal at the zone center, consistent with the notion that Ta$_2$NiSe$_5$ is a direct band gap system \cite{Wakisaka2009, Kaneko2012, Lu2017}.

Notably, as shown in Fig.~\ref{fig3}(b), while the onset of the RIXS excitations are identical, the leading edges and the high energy excitations exhibit notable differences between the Ta and Ni $L_3$-edges. The subtle differences between the two data sets also manifest in the momentum dependent gap maps shown in Fig.~\ref{fig3}(e) and (f). While both maps exhibit an apparent momentum dependence with a cusp where the gap size reaches its maximum and the gap edge is most extended, the momentum position of the cusp at the Ta $L_3$-edge is at $h \sim 0.25~r.l.u.$ from the zone center, whereas the cusp at the Ni $L_3$-edge is at $h \sim 0.3~r.l.u.$. We note that the subtle differences in the particle-hole continuum measured by the Ta and Ni $L$-edges are expected, as RIXS is element-specific and the measured particle-hole excitations should be weighted by the associated orbital characters of the corresponding element in the electronic structure. In other words, the RIXS particle-hole excitations measured at the Ta- and Ni-$L_3$-edges should be representative of the element-specific JDOS of Ta $5d$ and Ni $3d$ orbital characters, respectively. 

To further confirm that the RIXS maps reflect the element-specific JDOS, we performed element-specific JDOS calculations, using a band structure derived from a density functional theory (DFT) with the Perdew-Burke-Ernzerhof (PBE)-functional. The DFT bands, as well as the corresponding partial density of states of the Ta $5d$ (blue), Ni $3d$ (red), and Se $4p$ (green) orbitals, are plotted in Fig.~\ref{fig4}(a) and (b) for the monoclinic and orthorhombic structure, respectively. It is well-known that DFT with standard functionals like PBE routinely yields underestimated gap sizes. Therefore, we manually adjusted the band gap to match the experimental value in the low temperature monoclinic DFT calculation \cite{Wakisaka2009, Wakisaka2012, Kaneko2012, Fukutani2019}. The DFT calculations were done using the Vienna Ab initio Simulation Package (VASP) \cite{Kresse1993} for primitive cell structures with 18 $\times$ 18 $\times$ 6 Monkhorst-Pack k-point grids for both phases. The non-self-consistency calculations were then performed to obtain the orbital-resolved band structures for 100 $\times$ 10 $\times$ 10 k-point grids, with 100 k-point sampling along the chain direction. Element-specific JDOS were then calculated for Ta 5$d$ and Ni 3$d$. As shown in Fig.~\ref{fig4}(c) and (d), at low temperatures, the band gap in the Ta-specific JDOS and Ni-specific JDOS exhibits a momentum dependence which agrees with our experimental RIXS data. Furthermore, the different cusp momentum positions are reproduced in the Ta- and Ni-RIXS maps. However, while the gross features of the momentum dependent gap structure show good agreement, the JDOS intensity above the gap energy differs notably from that of the RIXS spectrum. We remark that while RIXS is sensitive to charge excitations, the cross-section is modulated additionally by the RIXS matrix elements, associated with particular light polarization and orbital character\cite{Jia2016}. Nevertheless, the close resemblance of the band gap and its momentum dependence between Ta and Ni $L_3$-edge RIXS data indicates that the Ta and Ni $d$-orbitals are well hybridized in the occupied valence and unoccupied conduction bands, consistent with the expected hybridizations in an EI state.

Next, we discuss the temperature dependence of the RIXS particle-hole excitation across the phase transition. Figure~\ref{fig5}(a) shows elastic peak subtracted Ta $L_3$-edge RIXS spectra at the zone center taken at temperatures across the phase transition. To quantitatively analyze the evolution of the RIXS gap edge, we fit the spectrum to an error function, which is the convolution between a step function and a normalized Gaussian function. The functional form of the error function is given by,
$$\frac{A}{2}[1+erf(\frac{2\sqrt{\log(2)}(\omega-\omega_0)}{\Gamma})]$$
where $\omega$ is the energy loss, $A$ and $\omega_0$ determine the height and the mid-point position of the step function, and $\Gamma$ represents the full-width-half-maximum of the normalized Gaussian function. As a reference, the fittings are also superimposed in Fig.~\ref{fig5}(a). We also define the energy position of 10 $\%$ step height as the onset of the gap edge and $\Gamma$ as the gap width. 

The extracted gap mid-point $\omega_0$, onset and width $\Gamma$ as a function of temperature are summarized in Fig.~\ref{fig5}(b). At 25 K, the RIXS gap leading edge is steep and the gap size is well defined by the onset energy. From 25 K to $T_c$, both the gap mid-point energy and the gap onset energy decrease, indicating the reduction of the gap size as the temperature increases. Meanwhile, the RIXS gap width increases by more than 3-fold with increasing temperature up to $T_c$, suggesting that electronic structure broadening also plays a role in the gap closing at high temperatures. These experimental findings are qualitatively consistent with the evolution of the gap measured by angle-resolved photoemission experiments \cite{Wakisaka2012, Tang2020}, in which the energy gap decreases, accompanied by a significant broadening of the quasi-particle peak approaching $T_c$.  Note that the gap measured by ARPES is the energy difference between the valence band and the Fermi energy, whereas the RIXS gap directly measures the full gap between of the valence and conduction bands. Above $T_c$, the low energy part of the spectrum only shows a slowly increasing slope without resolvable gap. Overall, the temperature dependence of RIXS gap evolves smoothly without any abrupt change across $T_c$, consistent with the order parameter evolution of a second-order phase transition.

Finally, we discuss the momentum-dependent Ta $L_3$-edge RIXS intensity map in the high temperature phase above $T_c$. As shown in Fig.~\ref{fig5}(c) and (d), in addition to the gap closing at the zone center, the broad gap edge at larger $q_{//}$ indicates that the corresponding particle-hole excitation involves the electronic states near the Fermi energy, as they exhibit dramatic broadening across the phase transition. Moreover, the overall momentum-dependence away from the zone center can still be recognized and is similar to that in the low temperature monoclinic phase. The RIXS gap structure is found to agree well with the calculated Ta-specific JDOS map (See Fig.~\ref{fig4}(e)) using the DFT bands calculated from the high temperature phase, implying that not only the overall energy-momentum dispersion, but also the Ta-character in the electronic band structure are well captured by the DFT calculations. We also remark that the similar momentum-resolved structure of the RIXS map across the phase transition implies a lack of dramatic variation of the associated elemental character or significant change in the Ta-Ni hybridization across $T_c$. This correlates with the small structural distortion across the phase transition in Ta$_2$NiSe$_5$.

\section*{Discussion}
In a simplified picture of Ta$_2$NiSe$_5$ in the high temperature phase, one usually associates the unoccupied conduction and occupied valence bands near the Fermi energy with pure Ta 5$d$ and Ni 3$d$ orbital characters, respectively. Above $T_c$, the hybridization between the bands with Ta and Ni characters is argued to be forbidden due to the protection from a mirror symmetry in the crystal structure \cite{Lu2017, Watson2020}. However, our results of element-specific RIXS particle-hole excitation and JDOS calculation indicate that the Ta and Ni orbital characters in the conduction and valence bands are already hybridized even above $T_c$. We also note that Ta and Ni atoms in Ta$_2$NiSe$_5$ are not located precisely on the same plane, which may cast some doubt on the validity of a mirror-symmetry argument that prevents Ta and Ni hybridization. These suggest that a more complex model might be needed for a more accurate account of the electronic structure in Ta$_2$NiSe$_5$. 

An important observation of our work is that the RIXS particle-hole excitations across the phase transition can be well described by the element-resolved JDOS calculations deduced from the DFT bands with corresponding crystal structures. This agreement indicates that the electronic structures, including the Ta $5d$ and Ni $3d$ hybridization, are essentially set by the corresponding lattice structure. This implies that additional Ta-Ni hybridization effect due to the putative EI transition might only play a minor role. In addition, despite that RIXS should be able to detect the electronic collective modes of the exciton condensation, no sharp peak or other structure that could be associated with a collective mode is found between the elastic peak and the rising edge of the RIXS spectrum within the energy resolution of our RIXS measurement at either the Ta or Ni $L_3$-edges (See Fig.~\ref{fig2}(e) and (f)). In other words, we do not find any signature of such electronic collective modes in the energy scale comparable with the low energy gap in the particle-hole excitations. This observation appears to support the argument that the putative EI transition, if it indeed exists, might be a relatively minor effect on top of the dominant structural transition. In such a scenario, the collective modes would likely lie at energies lower than our instrument resolution and are strongly mixed with the underlying lattice modes, consistent with the modes observed in Raman \cite{Kim2020} and time-resolved pump-probe experiments \cite{Mor2017, Werdehauseneaap8652}.  

In summary, we mapped out RIXS particle-hole excitations at both the Ta- and Ni $L_3$-edges, which reveal a momentum dependent gap in the spectra. Importantly, the gap structure agrees well with the Ta- and Ni-specific JDOS deduced from DFT band structures with corresponding lattice structures, providing information about the Ta-Ni hybridization in the electronic structure across the phase transition. Upon increasing the temperature, while the gap near the zone center closes, the Ta-character in the band structure does not exhibit dramatic changes across $T_c$, indicating that the Ta-Ni hybridization is essentially dictated by the lattice structure and agrees well with the DFT calculations. Our results might lend further support to the notion that the gap opening in Ta$_2$NiSe$_5$ is likely driven by the structural transition. We note that our results do not exclude a minor contribution from the EI state, whose proof of existence requires further scrutiny.

\section*{Acknowledgement}
This work is supported by the U.S. Department of Energy (DOE), Office of Science, Basic Energy Sciences, Materials Sciences and Engineering Division, under contract DE-AC02-76SF00515. We acknowledge the Advanced Photon Source, Argonne National Laboratory for providing beam time at the ID27-RIXS beamline under contract No. DE-AC02-06CH11357. We also acknowledge the Diamond Light Source, UK for providing beam time at the I21-RIXS beamline.

%

\end{document}